# Global near-real-time daily emissions of atmospheric pollutants from power plants

Tao Li[1], Lixing Wang[1], Biqing Zhu[2], and Zhu Liu[1,3,4]

[1]Department of Earth System Science, Tsinghua University, Beijing 100084, China

[2]International Institute for Applied Systems Analysis (IIASA), Laxenburg, Austria

[3]Department of Earth System Science, Institute for Carbon Neutrality, State Key Laboratory of Hydroscience and Engineering, Tsinghua University, Beijing 10084, China

[4]Yellow River Laboratory (Henan), Zhengzhou 450046, China

*Correspondence to*: Zhu Liu (zhuliu@tsinghua.edu.cn)

**Abstract.** The power sector is a major source of fossil fuel consumption and air pollutant emissions. High-spatiotemporal-resolution emission accounting is therefore essential for the precise formulation of mitigation policies and the strengthening of air quality management. However, existing publicly available emission inventories remain limited in timeliness and spatiotemporal resolution because of constraints in data availability and completeness. Here, we construct a global, plant-level, daily, multi-pollutant emission database for the power sector by integrating nearly 3 million hourly-to-daily near-real-time multi-energy power generation records from 57 countries (accounting for approximately 81% of global fossil-fuel-based electricity generation) with fundamental information on more than 10,000 power plants worldwide, including plant location and installed capacity. The resulting dataset substantially improves the timeliness and granularity of global power-sector emission accounting. Results show that global power-plant emissions of most pollutants increased during 2019–2025, with 2025 daily mean emissions reaching 0.274 kt/d for BC, 45.1 kt/d for CO, 0.418 kt/d for $NH_3$, 52.2 kt/d for $NO_x$, 3.01 kt/d for NMVOC, 0.418 kt/d for OC, 6.76 kt/d for $PM_{10}$, 5.11 kt/d for $PM_{2.5}$, and 78.5 kt/d for $SO_2$. Relative to 2019, NMVOC exhibited the largest increase, whereas $SO_2$ was the only pollutant showing a net decline. Coal remained the dominant source of sulfur-, nitrogen-, and particulate-related emissions, while gas and biomass contributed more prominently to carbonaceous species and reduced nitrogen. The dataset also captures pronounced seasonal, regional, and short-term variations in emissions. Compared with the benchmark EDGAR inventory over 2019-2022, our estimates show strong agreement, with Pearson correlation coefficients ranging from 0.92 to 0.99 and mean relative deviations ranging from 8.8% to 28.1%. This near-real-time, high-resolution dataset provides robust scientific support for coordinated air pollution control and carbon mitigation in the power sector and offers an important data foundation for improving global emission monitoring and satellite-based inversion systems.

## 1. Introduction

As a primary contributor to fossil fuel consumption as well as air pollutant emissions, the power sector exerts a decisive influence on regional air quality improvement (Tong et al., 2018; Li et al., 2025). Accurately characterizing the near-real-time evolution of power-sector pollutant has therefore become a central research focus and an urgent scientific priority within the research community (Soulie et al., 2024; Zheng et al., 2021). The development and application of near-real-time, high-spatiotemporal-resolution power emission datasets hold practical significance for global regional environmental management (Soulie et al., 2024; Tong et al., 2021; Zheng et al., 2021).

On the one hand, such datasets provide critical support for the formulation and optimization of adaptive mitigation policies (Soulie et al., 2024; Zheng et al., 2021; Wu et al., 2022; Li et al., 2023; Liu et al., 2024a). As electricity consumption structures continuously evolve alongside industrial upgrading, the integration of renewable energy and the complementary operation of fossil-fuel-based generation introduce substantial variability into emission patterns (Zhao et al., 2023). These fluctuations may induce short-term shifts in emission characteristics (Li et al., 2026b). Near-real-time emission data enable precise tracking of these dynamic processes, facilitating the rapid identification of high-emission periods and hotspot regions (Li et al., 2025; Dou et al., 2021). This capability supports the design of differentiated mitigation strategies, enhances the timeliness and accuracy of pollution control efforts, and overcomes the inherent lag associated with conventional regulatory approaches (Liu et al., 2020; Ke et al., 2023; Liu et al., 2022c; Zheng et al., 2021).

On the other hand, near-real-time high-resolution emission data provide reliable ground-truth information for the validation and extrapolation of satellite-based emission inversion techniques. With advances in satellite remote sensing, instruments such as TROPOspheric Monitoring Instrument (TROPOMI) (Veefkind et al., 2012) and Geostationary Environment

Monitoring Spectrometer (GEMS) (Lee et al., 2024) have enabled large-scale and routine monitoring of major air pollutants. However, the accuracy of satellite inversion products depends heavily on high-quality ground observations or robust emission inventories for baseline constraint and calibration (Lin et al., 2023; Zheng et al., 2020; Lin et al., 2024; He et al., 2022). Near-real-time, high-spatiotemporal-resolution power emission data can effectively bridge spatiotemporal gaps in ground-based monitoring, support parameter optimization and validation of inversion models, and promote the extension of satellite-based monitoring from coarse regional scales to finer spatial and temporal resolutions, thereby enhancing the overall reliability of global and regional emission surveillance systems (Lin et al., 2023; Zheng et al., 2021; Zheng et al., 2020).

Although mainstream global emission inventories have provided essential data for power-sector emission research, their structural limitations constrain their applicability to near-real-time and high-resolution analyses. Classical inventories such as Emissions Database for Global Atmospheric Research (EDGAR) (Crippa et al., 2024; Janssens-Maenhout et al., 2019) and Community Emissions Data System (CEDS) (Hoesly et al., 2018) offer long-term historical emission records and extensive spatial coverage. EDGAR provides global multi-pollutant emission data from 1970 onward, while CEDS extends back to 1750, representing one of the longest available emission time series. Nevertheless, both inventories suffer from substantial time lags, typically updating data with a delay of two to three years, which limits their capacity to capture recent emission dynamics. To address this limitation, products such as Copernicus Atmosphere Monitoring Service (CAMS-GLOB-ANT) (Soulie et al., 2024) have introduced near-real-time emission datasets designed to support air quality forecasting and reanalysis. However, these datasets generally rely on historical trends and predefined temporal profiles, and may not yet fully incorporate near-real-time activity data such as actual power generation and fuel consumption structures. As a result, they might not completely capture the detailed spatiotemporal variability of power-sector emissions, which could affect their applicability for precision mitigation policy design and satellite inversion validation, particularly in the context of rapidly evolving energy systems.

In this study, we utilize near-real-time power generation data covering 57 countries, comprising nearly 3 million hourly-to-daily records across multiple energy sources (e.g., coal, gas, oil, and biomass), together with detailed information on the geographic location and installed capacity of over 10,000 power plants worldwide. Based on these data, we construct a global, plant-level, daily emission database encompassing multiple air pollutants, thereby substantially enhancing the spatiotemporal resolution of power-sector emission accounting. The resulting near-real-time, high-resolution emission dataset advances the granularity and responsiveness of power emission monitoring and provides critical scientific support for coordinated air pollution control in the power sector, as well as for the continuous improvement of global emission monitoring frameworks.

## 2. Methods

Air pollutant emissions are generally estimated using activity data and emission factors, as expressed in the following equation (Janssens-Maenhout et al., 2019; Tong et al., 2021):

$$E_{c,y,s,i} = AD_{c,y,i} \times EF_{c,y,s,i} \tag{1}$$

where $c$, $y$, $s$, and $i$ represent country, year, emission specie, and fuel type, respectively; $E$ denotes emissions (kg), $AD$ represents activity data (kg for solid- and liquid-fueled units and $m^3$ for gas-fired units), and $EF$ denotes emission factors. Emission factors quantify the mass of pollutants emitted per unit of activity and depend on fuel characteristics (e.g., sulfur, carbon, and ash content), combustion technology, and the implementation of emission control measures. These factors are derived from authoritative international and national databases, including the International Energy Agency (IEA) and the United States Geological Survey (USGS), as well as from the scientific literature (Tong et al., 2018; Huang et al., 2017; Zhong et al., 2020; Huang et al., 2014; Shan et al., 2025). Activity data document fuel consumption, and are typically compiled in national statistical yearbooks or by the IEA (IEA, 2025). However, such data are generally released on an annual basis with a time lag of approximately two years, limiting their applicability for estimating near-real-time daily emissions. Recent advances in low-latency, high-frequency power generation reporting provide a viable pathway for estimating emissions in near real time (Zhu et al., 2023; Liu et al., 2020; Ke et al., 2023; Huo et al., 2022; Liu et al., 2022a; Liu et al., 2024b; Liu et al., 2023; Deng et al., 2025; Liu et al., 2022b; Zheng et al., 2020; Zheng et al., 2021). These studies compile power generation data from various publicly available sources, including European Network of Transmission System Operators for Electricity (ENTSO-E) for European countries (Hirth et al., 2018), Eskom for South Africa (Eskom, 2026),

OpenNEM for Australia (OpenNEM, 2026), and Organization for Cross-regional Coordination of Transmission Operators (OCCTO) for Japan (OCCTO, 2026). In this study, near three million hourly to daily generation records, beginning in 2019, were collected from these public sources and harmonized into daily power generation data, hereafter referred to as CarbonMonitor-Power. Detailed descriptions of the data acquisition and processing procedures are provided in our previous studies (Zhu et al., 2023; Li et al., 2026b).

However, in most publicly available datasets, power generation is not reported for individual fuel types but rather for aggregated categories. For example, anthracite, lignite, and sub-bituminous coal are collectively classified under the "coal" category. To utilize such datasets for estimating air pollutant emissions, Eq. (1) is reformulated as:

$$E_{c,y,s,m,j} = PC_{c,y,m,j} \times EF_{c,y,s,m} \tag{2}$$

where $m$ denotes the fuel category (i.e., coal, gas, oil, and biomass), representing an aggregation of specific fuel types; $j$ indicates $j$th day of year $y$; $PC$ represents electricity generation (GWh) compiled in CarbonMonitor-Power; $EF$ denotes the corresponding emission factor expressed in kg/GWh, derived from literature (Crippa et al., 2024; Janssens-Maenhout et al., 2019; Tong et al., 2018). Table S1 presents emission factors (kg/GWh) for coal-, gas-, oil-, and biomass-fired power generation in 2019, highlighting clear differences across fuel types. Coal- and oil-fired generation exhibit the highest emission factors, particularly for $SO_2$ and $NOx$, whereas gas-fired plants show consistently lower emissions across most pollutants. Biomass-fired generation displays the greatest variability and relatively high emissions of CO, NMVOC, and particulate matter, reflecting heterogeneity in fuel characteristics and combustion conditions.

**Table 1**. Countries and regions included in the near-real-time power generation dataset used in this study.

| Continent | Available countries/regions |
| --- | --- |
| Europe | Austria, Belgium, Bosnia & Herz, Bulgaria, Croatia, Czech Republic, Denmark, Estonia, Finland, France, Germany, Greece, Hungary, Italy, Kosovo, Latvia, Lithuania, Luxembourg, Moldova, Montenegro, Netherlands, North Macedonia, Norway, Poland, Portugal, Romania, Russia, Serbia Slovakia, Slovenia, Spain, Sweden, Switzerland, United Kingdom, |
| Asia | Bangladesh, China, Georgia, India, Japan, South Korea, Thailand, Turkey |

| Africa | South Africa, Nigeria |
| Americas | Argentina, Bolivia, Brazil, Chile, Costa Rica, Dominican Republic, El Salvador, Mexico, Peru, United States, Uruguay |
| Oceania | Australia, New Zealand |

Excluding countries with an entire year of missing data, CarbonMonitor-Power covers 57 countries (Table 1), accounting for approximately 81% of global fossil-fuel-based electricity generation, according to the IEA (IEA, 2025). For countries not included, annual power generation data from the IEA were extrapolated and temporally disaggregated to derive daily estimates. As the latest IEA data are available up to 2023, continent-level growth rates derived from CarbonMonitor-Power were applied to extend national generation to 2024–2026:

$$r_{n,y,m} = \frac{PC_{n,y+1,m}}{PC_{n,y,m}} \tag{3}$$

$$PE_{c,y+1,m} = r_{n,y,m} \times PI_{c,y,m} (c \in n) \tag{4}$$

where $r$ represent growth rate, $n$ denotes continent, and $PI$ represent power generation from the IEA, respectively.

Annual generation was then disaggregated to daily resolution using temporal profiles from CarbonMonitor-Power:

$$PE_{c,y,m,j} = PE_{c,y,m} \times \frac{PC_{n,y,m,j}}{PC_{n,y,m}} (c \in n) \tag{5}$$

where $j$ represent $j$th day of year $y$. Daily emissions were estimated by combining activity data with emission factors:

$$E_{c,y,s,m,j} = PE_{c,y,m,j} \times EF_{c,y,s,m} \tag{7}$$

National near-real-time emissions estimated using Eqs. (2) and (7) were further allocated to the power plant level based on installed capacity:

$$E_{c,y,s,m,j,k} = \frac{Cap_{c,m,k}}{\sum_k Cap_{c,m,k}} \times E_{c,y,s,m,j} \tag{8}$$

where $k$ indicates the $k$th power plant of country $c$, and $Cap$ denotes the capacity of power plant. Following the methodology of the EDGAR, this study integrates multiple complementary global and national power plant datasets to achieve comprehensive coverage across fuel types (Crippa et al., 2024). Specifically, coal-fired power plants were obtained from the Global Coal Plant Tracker (GEM, 2025a), while gas-fired units were primarily derived from the Global Oil and Gas Plant Tracker (GEM, 2025b). Additional oil- and biomass-fired plants were supplemented using the Global Power Plant

Database (Byers et al., 2018), and a dedicated national dataset was incorporated to improve coverage and accuracy for the United States (EIA, 2022).

To ensure temporal consistency with the selected base year (e.g., 2020), a rigorous filtering procedure was applied to identify plants that were active or relevant during that year. Plants were restricted to those classified as operating, mothballed, or retired, while those labeled as proposed or under construction were excluded. Operating plants commissioned after the base year were removed. Mothballed plants were evaluated using a 40-year lifetime assumption and excluded if their commissioning year plus 40 years preceded the base year. Retired plants were excluded if their retirement year was earlier than the base year; for plants lacking retirement year information, the same lifetime assumption was applied. For datasets without explicit status information, all plants were initially treated as operating, and the 40-year lifetime criterion was used to exclude units likely retired prior to the base year. This filtering strategy ensures a consistent temporal boundary across heterogeneous data sources and enhances the robustness of plant-level emission estimates.

## 3. Results

### 3.1 Global patterns of power plant emissions

Figures 1 and S1-S8 illustrate the spatial distribution of global power plants and their corresponding emissions across four fuel categories: coal, oil, gas, and biomass. Coal-fired power plants exhibit the highest emission intensities and are densely clustered in China, India, eastern North America, and Western Europe, reflecting the dominant role of coal in regional power generation. Oil-fired power plants are more geographically dispersed, with notable concentrations in North America, South America, and South Asia. Gas-fired plants are primarily concentrated in North America, Europe, and South Asia, consistent with the distribution of natural gas infrastructure. Biomass-fired power plants show the most limited spatial extent, with relatively higher emissions concentrated in Europe and parts of North America, in line with regional biomass utilization policies.

From 2019 to 2025, global emissions of most pollutants exhibited an overall upward tendency, although the interannual variations differed among species (Fig. 2). By 2025, the global daily mean emissions reached 0.274 kt/d for BC, 45.1 kt/d for CO, 0.418 kt/d for NH$_3$, 52.2 kt/d for NOx, 3.01 kt/d for NMVOC, 0.418 kt/d for OC, 6.76 kt/d for PM$_{10}$, 5.11 kt/d for PM$_{2.5}$, and 78.5 kt/d for SO$_2$. Relative to 2019, BC, CO, NH$_3$, NOx, NMVOC, OC, PM$_{10}$, and PM$_{2.5}$ increased by 17.1%, 20.9%, 20.5%, 8.3%, 29.7%, 22.6%, 22.2%, and 22.0%, respectively, whereas SO$_2$ decreased by 3.0%. Compared with 2024, emissions in 2025 increased slightly for CO (+0.3 kt/d), OC (+0.003 kt/d), and PM$_{10}$ (+0.03 kt/d), remained nearly unchanged for NMVOC and PM$_{2.5}$, but declined for BC (-0.003 kt/d), NH$_3$ (-0.002 kt/d), NOx (-1.2 kt/d), and especially SO$_2$ (-5.4 kt/d). Overall, NMVOC showed the largest relative increase during 2019-2025, while SO$_2$ was the only pollutant with a net decline.

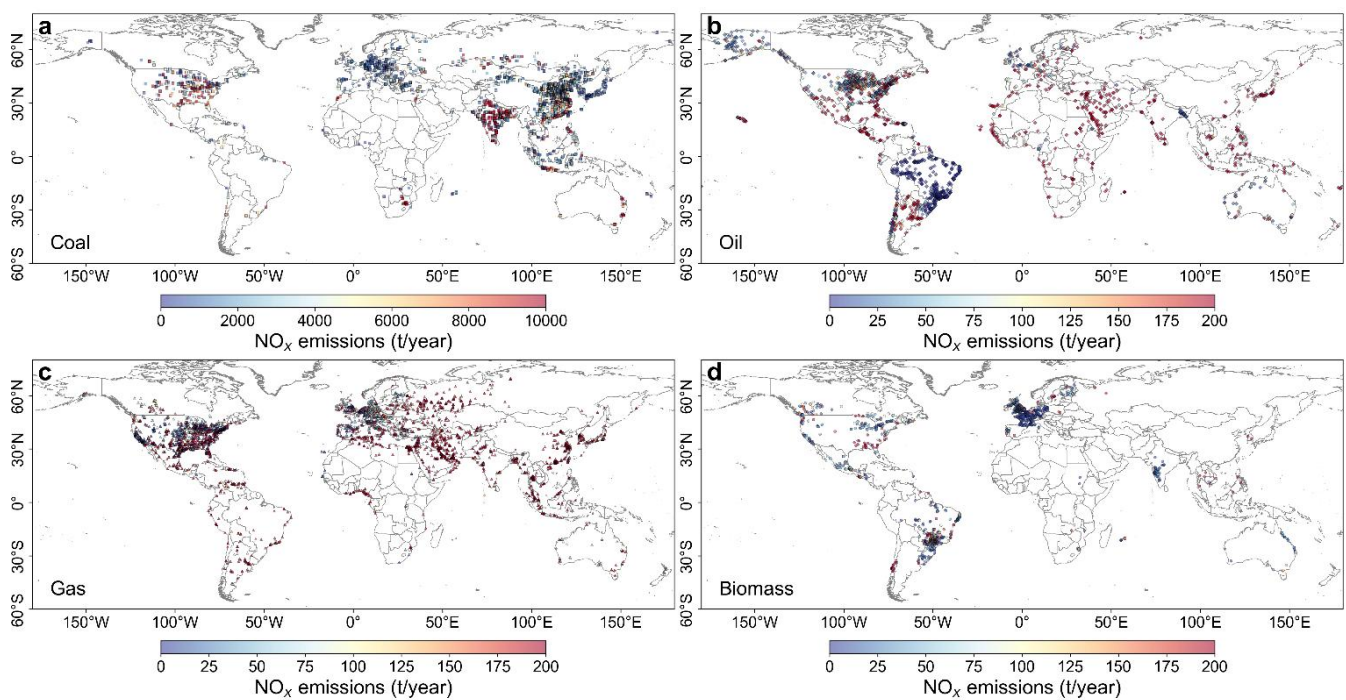

**Figure 1.** Spatial distribution of power plants and their associated NOx emissions for (a) coal, (b) oil, (c) gas, and (d) biomass.

The sectoral structure in 2019 exhibited marked pollutant dependence. For the key pollutants SO$_2$, NOx, and PM$_{2.5}$, coal was the dominant source, with daily mean emissions of 58.5, 34.0, and 2.65 kt/d, respectively. Oil was also an important contributor to $SO_2$ and NOx, with emissions of 20.8 and 6.68 kt/d, respectively, while biomass made a substantial contribution to $PM_{2.5}$ (1.23 kt/d). For the other pollutants, coal remained the leading source of BC and $PM_{10}$ (0.140 and 2.80 kt/d), gas dominated CO and NMVOC emissions (24.0 and 0.816 kt/d), and biomass was the principal source of $NH_3$ and OC (0.265 and 0.167 kt/d). These results indicate that coal mainly controlled sulfur-, nitrogen-, and particulate-related emissions, whereas gas and biomass played more prominent roles in carbonaceous and reduced nitrogen pollutants.

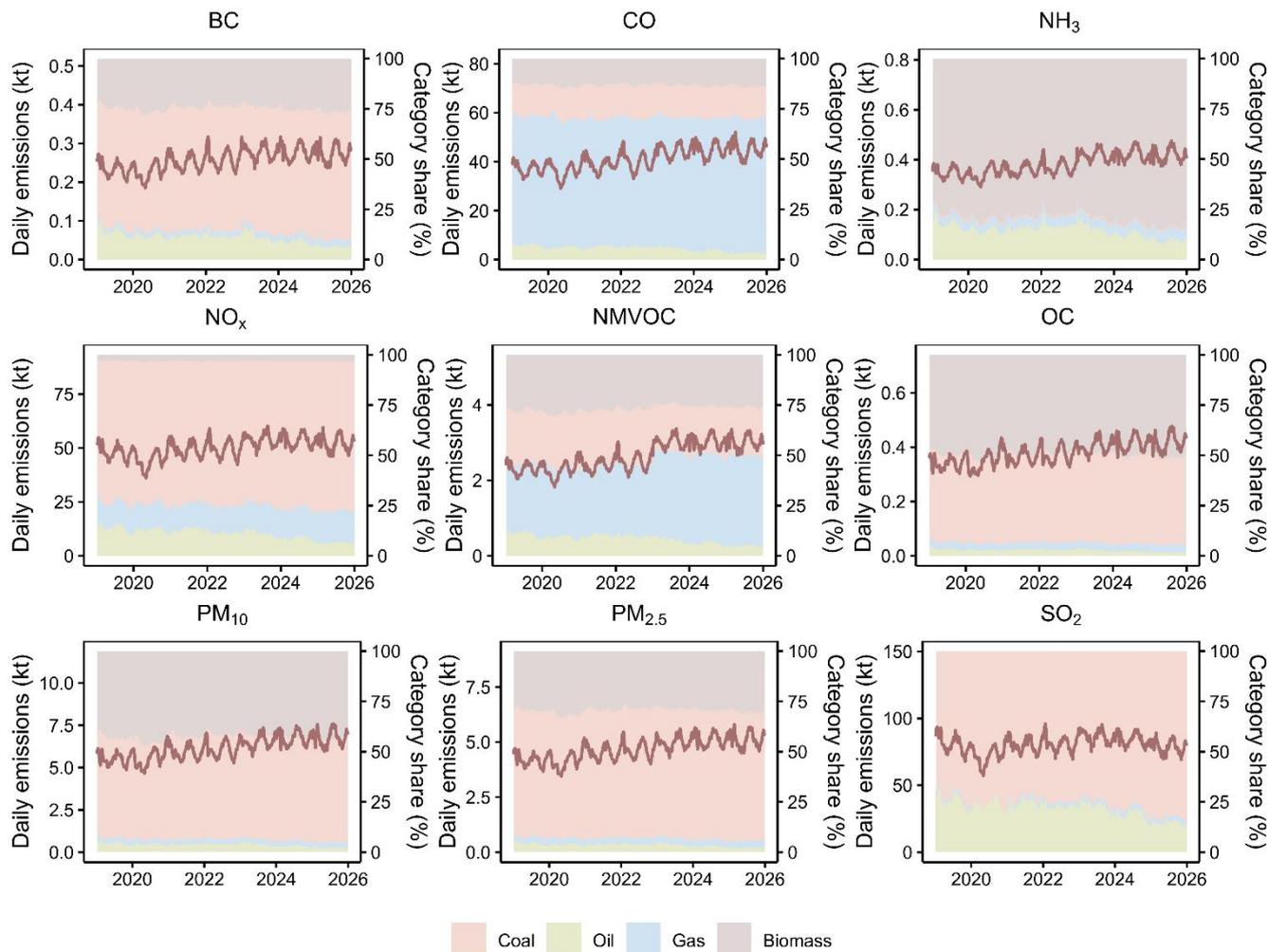

**Figure 2.** Global daily total emissions and fuel-category shares of power-sector emissions by pollutant.

From 2019 to 2025, sector-specific emissions evolved in distinctly different ways. Coal-related emissions generally increased in absolute terms, especially for $SO_2$, NOx, $PM_{10}$, and $PM_{2.5}$, although most coal-related pollutants showed a slight decline from 2024 to 2025. In contrast, oil-related emissions decreased systematically for all pollutants over 2019–2025 and continued to decline in 2025 relative to 2024, indicating a broad weakening of oil-related emission contributions. By comparison, gas-related emissions increased for all pollutants during 2019–2025, with particularly strong increases in BC (+73.7%), NMVOC (+66.7%), $NH_3$ (+51.6%), and $SO_2$ (+47.2%) relative to 2019; most gas-related pollutants also continued to increase in 2025 compared with 2024. Biomass-related emissions likewise showed a consistent upward tendency, with increases across all pollutants during 2019–2025 and further increases in 2025, especially for $NH_3$, OC, $PM_{10}$, and $PM_{2.5}$.

The monthly evolution of emissions displayed a clear seasonal cycle (Fig. 3). In general, emissions were relatively low in boreal spring, especially in April, and relatively high in winter and late summer. For most pollutants, the annual maximum occurred in August, including BC (8.74 kt), $NH_3$ (12.9 kt), NOx (1690 kt), NMVOC (88.9 kt), OC (12.9 kt), $PM_{10}$ (207 kt), and $PM_{2.5}$ (156 kt), whereas CO and $SO_2$ peaked in January, reaching 1380 and 2690 kt, respectively. The annual minima generally occurred in April, with values of 7.06 kt for BC, 1120 kt for CO, 10.7 kt for $NH_3$, 1400 kt for NOx, 73.3 kt for NMVOC, 171 kt for $PM_{10}$, 130 kt for $PM_{2.5}$, and 2260 kt for $SO_2$; OC reached its minimum in February (10.4 kt). This seasonal pattern suggests that global pollutant emissions were characterized by a springtime trough and enhanced emissions during winter and/or late summer.

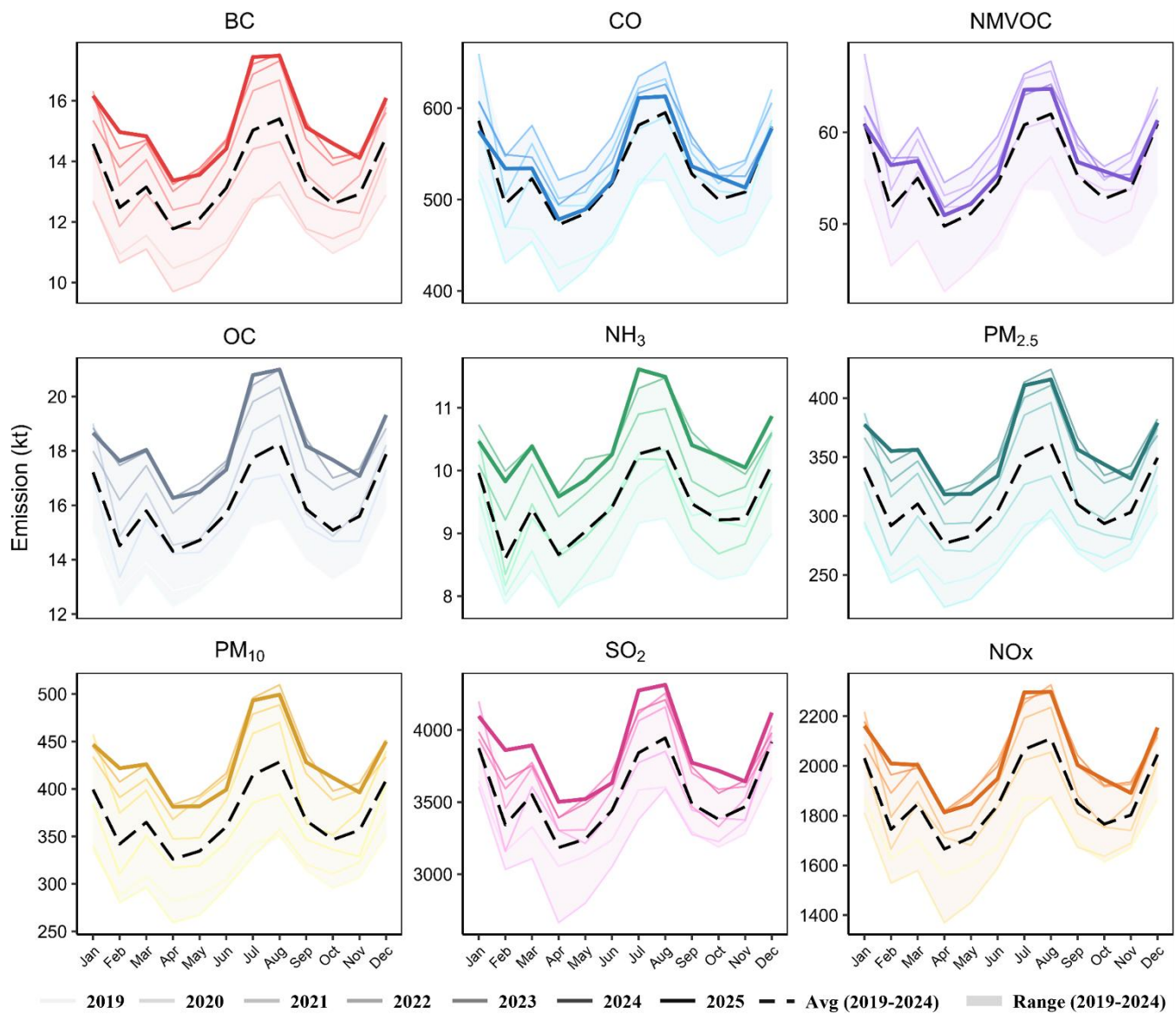

**Figure 3.** Global monthly total emissions from power plants by pollutant.

## 3.2 Regional patterns of power plant emissions

### 3.2.1 China

China remained one of the largest contributors to global power-plant emissions during the study period (Fig. 4). In 2019, the daily mean emissions of BC, CO, NH₃, NOx, NMVOC, OC, PM₁₀, PM₂.₅, and SO₂ were 0.0644, 5.19, 0.0884, 11.8, 0.517, 0.163, 1.97, 1.37, and 13.6 kt/d, respectively. These values accounted for 27.5%, 13.9%, 25.5%, 24.5%, 22.3%, 47.8%, 35.6%, 32.7%, and 16.8% of the corresponding global power-sector emissions, confirming China as a major global source of power-plant pollution.

Coal was the dominant emitting fuel in China, especially for NOx and SO₂. In 2019, coal contributed 56.0% of BC, 66.8% of CO, 1.89% of NH₃, 93.0% of NOx, 73.4% of NMVOC, 46.3% of OC, 39.9% of PM₁₀, 57.3% of PM₂.₅, and 97.8% of SO₂ emissions from the power sector. In absolute terms, coal-fired power plants emitted 0.0360, 3.46, 0.00167, 11.0, 0.379, 0.0757, 0.787, 0.787, and 13.3 kt/d of the above pollutants, respectively. By contrast, gas emissions were only $1.46{\times}10^{-4}$, 0.556, $6.94{\times}10^{-4}$, 0.398, 0.0180, $5.84{\times}10^{-4}$, 0.00584, 0.00584, and 0.00182 kt/d; oil emissions were $1.78{\times}10^{-4}$, 0.0245, $8.71{\times}10^{-4}$, 0.0692, 0.00249, $1.19{\times}10^{-4}$, 0.00245, 0.00174, and 0.245 kt/d; and biomass emissions were 0.0280, 1.14, 0.0851, 0.366, 0.117, 0.0870, 1.18, 0.580, and 0.0484 kt/d, respectively.

From 2019 to 2025, China's power-plant emissions continued to rise. Daily mean BC, CO, NH₃, NOx, NMVOC, OC, PM₁₀, PM₂.₅, and SO₂ increased by 28.9%, 31.6%, 42.5%, 9.3%, 28.2%, 31.9%, 33.5%, 29.2%, and 14.7%, respectively. This indicates that, despite continuous progress in pollution control and the rapid deployment of non-fossil power, growing electricity demand still drove an overall increase in emissions from the power sector. However, the relative dependence on coal weakened. Between 2019 and 2025, coal-related emissions increased more slowly than total emissions, by 17.5% (BC), 19.9% (CO), 18.6% (NH₃), 6.4% (NOx), 18.5% (NMVOC), 17.6% (OC), 17.4% (PM₁₀), 17.4% (PM₂.₅), and 15.0% (SO₂), respectively, and coal's contribution to total emissions declined from 56.0% to 51.0% for BC, 66.8% to 60.7% for CO, 1.89% to 1.57% for NH₃, 93.0% to 90.3% for NOx, 73.4% to 67.8% for NMVOC, 46.3% to 41.4% for OC, 39.9% to 35.1% for PM₁₀, and 57.3% to 52.3% for PM₂.₅, although SO₂ remained overwhelmingly coal-dominated (97.8% to 98.3%). In contrast, gas- and biomass-related emissions increased much faster. Gas-related emissions rose by 67.8%, 70.5%, 65.7%, 67.1%, 73.3%, 67.6%, 67.6%, 67.6%, and 67.6%, while biomass-related emissions increased by 43.6%, 50.0%, 43.4%, 43.4%, 53.8%, 43.7%, 43.2%, 43.6%, and 43.4%, respectively. These changes suggest an accelerated substitution toward cleaner fossil fuel and renewable biomass, although not sufficient to offset the increase in total emissions.

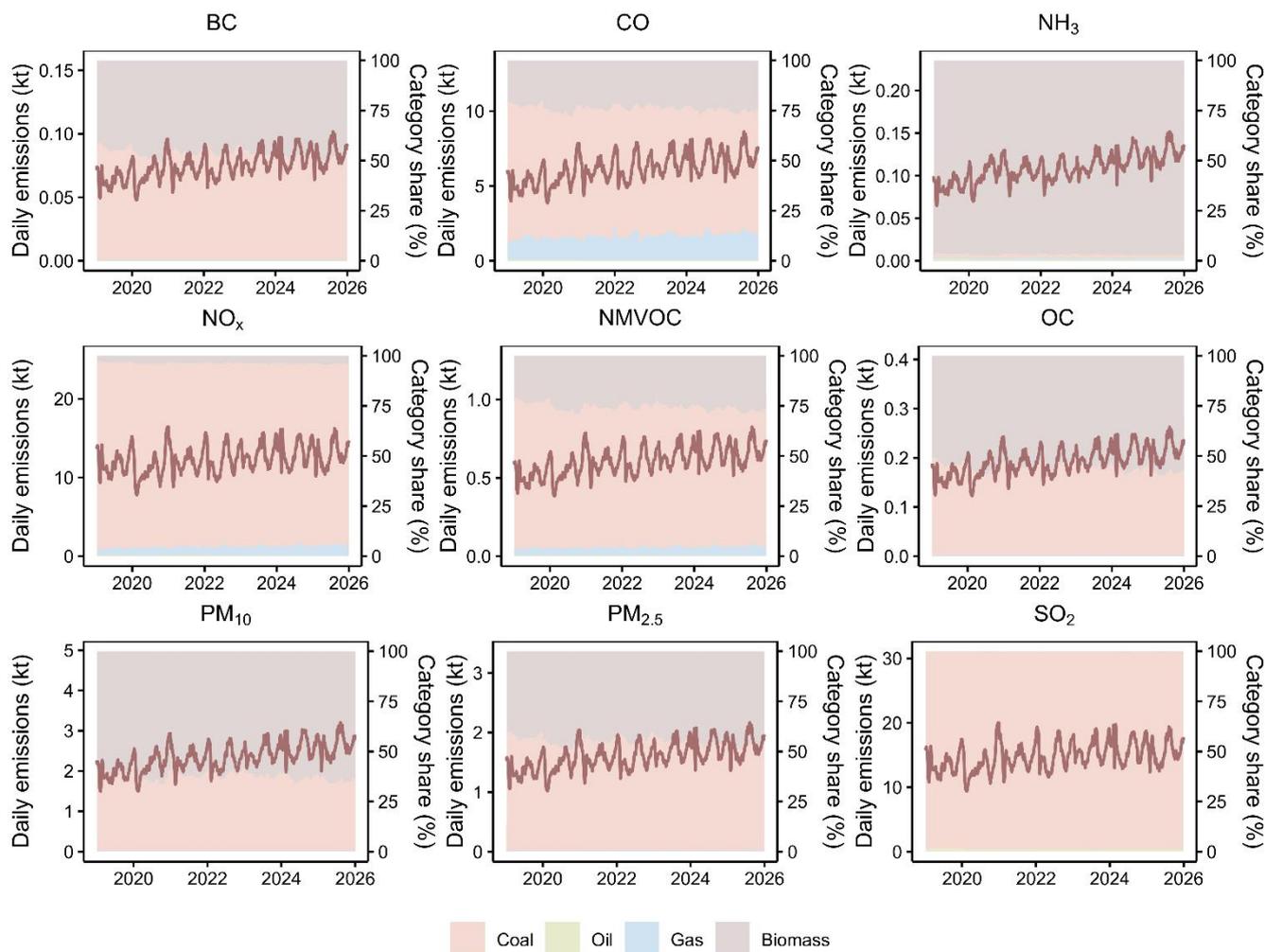

**Figure 4.** Daily total emissions and fuel-category shares of power-sector emissions in China by pollutant

China also exhibited a clear bimodal seasonal pattern in power-plant emissions, with major peaks in August and December and minima mainly in February and April (Fig. S9). CO, NH₃, OC, and PM₁₀ reached their annual maxima in August, at 223, 3.84, 6.82, and 83.0 kt/month, respectively, whereas BC, NOx, NMVOC, PM₂.₅, and SO₂ peaked in December, at 2.67, 454, 21.9, 56.9, and 548 kt/month, respectively. This bimodal structure likely reflects the combined effects of summer electricity demand and winter heating-related power generation.

### 3.2.2 India

Coal likewise dominated power-plant emissions in India and remained the main driver of their increase (Fig. 5). In 2019, coal-fired power plants emitted 0.0314, 0.865, $3.80 \times 10^{-4}$, 8.72, 0.0675, 0.0266, 0.725, 0.725, and 16.8 kt/d of BC, CO, $NH_3$, NOx, NMVOC, OC, $PM_{10}$, $PM_{2.5}$, and $SO_2$, respectively. These emissions accounted for 82.0%, 51.9%, 1.03%, 96.7%, 44.9%, 56.3%, 72.7%, 84.1%, and 99.3% of India's total power-sector emissions, indicating that coal overwhelmingly dominated the national power-related pollution burden. By comparison, gas, oil, and biomass emissions were much smaller. In 2019, gas emitted $2.47 \times 10^{-5}$, 0.132, $1.18 \times 10^{-4}$, 0.0705, 0.00844, $9.88 \times 10^{-5}$, $9.88 \times 10^{-4}$, $9.88 \times 10^{-4}$, and $3.11 \times 10^{-4}$ kt/d; oil emitted $3.01 \times 10^{-4}$, 0.0380, $3.42 \times 10^{-4}$, 0.0645, 0.00336, $6.79 \times 10^{-5}$, 0.00172, 0.00121, and 0.104 kt/d; and biomass emitted 0.00658, 0.630, 0.0362, 0.162, 0.0709, 0.0204, 0.269, 0.134, and 0.0206 kt/d, respectively.

From 2019 to 2024, India's power-plant emissions increased rapidly. BC, CO, $NH_3$, NOx, NMVOC, OC, $PM_{10}$, $PM_{2.5}$, and $SO_2$ changed by +28.2%, +18.7%, -11.6%, +40.8%, +15.3%, +16.3%, +27.5%, +33.6%, and +34.3%, respectively. The increase was almost entirely driven by coal-fired power generation. Over the same period, coal-related emissions increased by 36.9%, 42.2%, 41.3%, 42.2%, 46.7%, 39.1%, 42.1%, 42.1%, and 33.9%, whereas gas-related emissions declined by 6.5%, 9.1%, 6.8%, 7.1%, 18.0%, 6.4%, 6.5%, 6.5%, and 6.4%, and biomass-related emissions decreased by 13.7%, 10.6%, 13.0%, 11.1%, 11.8%, 13.2%, 13.8%, 13.4%, and 13.1%, respectively. These results indicate that India's post-pandemic growth in electricity demand was mainly met by coal-fired generation, further strengthening the coal dependence of the power sector and aggravating its associated air-pollution burden. From 2024 to 2025, however, India's power-sector emissions showed a slight decline. BC, CO, $NH_3$, NOx, NMVOC, OC, $PM_{10}$, $PM_{2.5}$, and $SO_2$ changed by -2.6%, -3.6%, +0.6%, -3.1%, -2.3%, -1.6%, -2.4%, -2.6%, and -3.1%, respectively. Nearly all pollutants decreased modestly in 2025, suggesting that emission growth may have started to slow, possibly reflecting the initial effects of pollution control and structural adjustment.

Unlike the bimodal seasonality in China, India showed relatively weak seasonality and an overall single-peak pattern (Fig. S10). BC, NOx, OC, $PM_{10}$, $PM_{2.5}$, and $SO_2$ all reached their annual maxima in March, at 1.41, 353, 1.65, 36.9, 32.9, and 637 kt/month, respectively. CO and NMVOC peaked in May, at 58.0 and 5.20 kt/month, whereas $NH_3$ peaked in July at 1.16 kt/month. All pollutants reached their minima in November, with BC, CO, NH$_3$, NOx, NMVOC, OC, PM$_{10}$, PM$_{2.5}$, and SO$_2$ decreasing to 1.23, 49.9, 0.976, 306, 4.49, 1.44, 32.2, 28.7, and 552 kt/month, respectively.

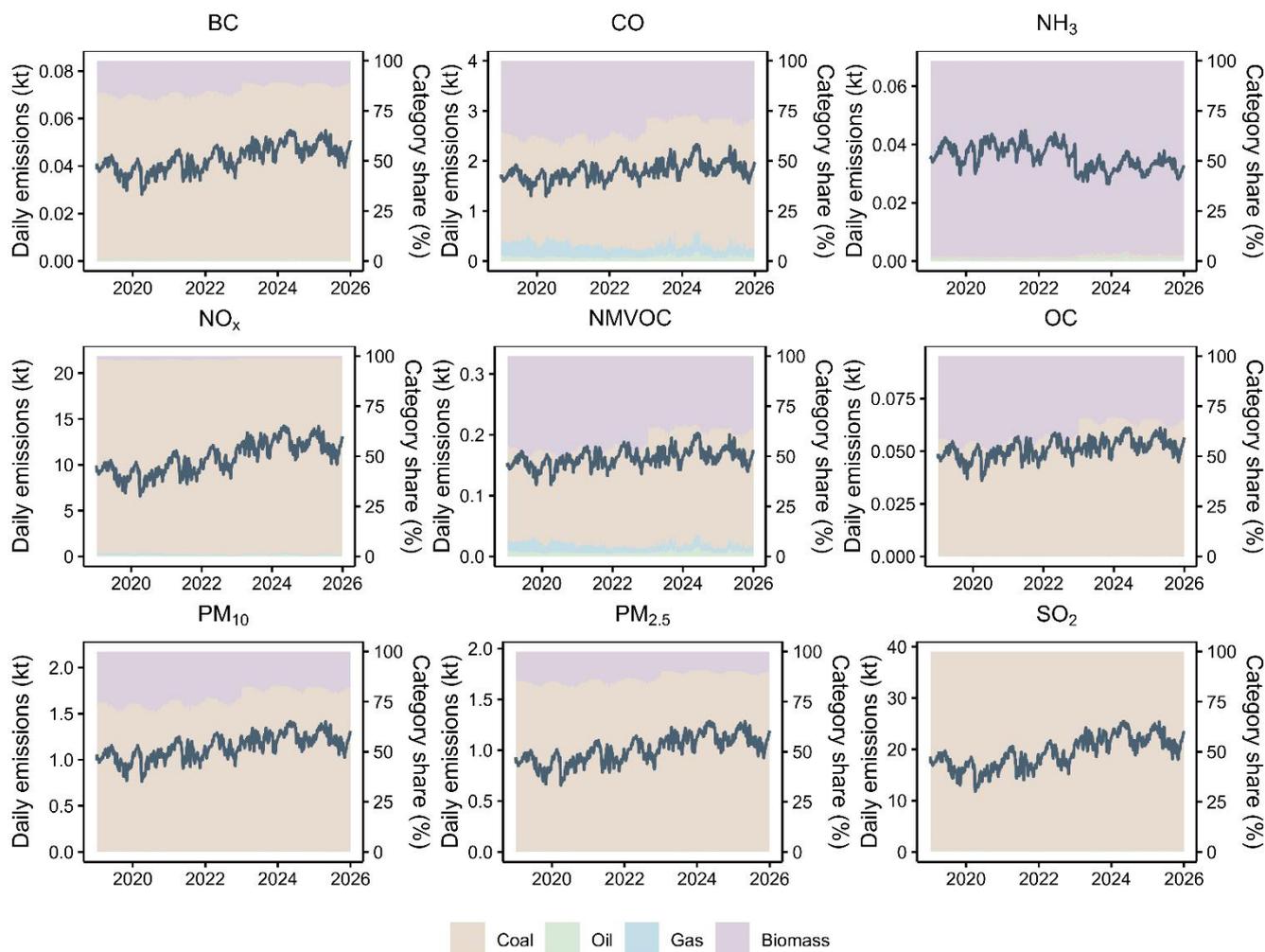

**Figure 5.** Daily total emissions and fuel-category shares of power-sector emissions in India by pollutant

### 3.2.3 United States

The United States exhibited relatively large but overall stable power-plant emissions during 2019–2025 (Fig. 6). In 2019, the daily mean emissions of BC, CO, NH₃, NOx, NMVOC, OC, PM₁₀, PM₂.₅, and SO₂ were 0.0335, 5.83, 0.0361, 5.13, 0.240, 0.0252, 0.567, 0.493, and 4.03 kt/d, respectively, accounting for 14.3%, 15.6%, 10.4%, 10.6%, 10.3%, 7.39%, 10.3%, 11.8%, and 4.98% of global power-sector emissions. From 2019 to 2025, BC, CO, NH₃, NMVOC, OC, PM₁₀, and PM₂.₅ increased by 13.1%, 12.3%, 9.4%, 9.2%, 7.5%, 10.6%, and 13.2%, respectively, whereas NOx and SO₂ declined by 4.7% and 30.8%. This indicates that U.S. power-sector emissions remained broadly stable in aggregate, but with a clear reduction in sulfur emissions and a modest rise in carbonaceous and particulate pollutants.

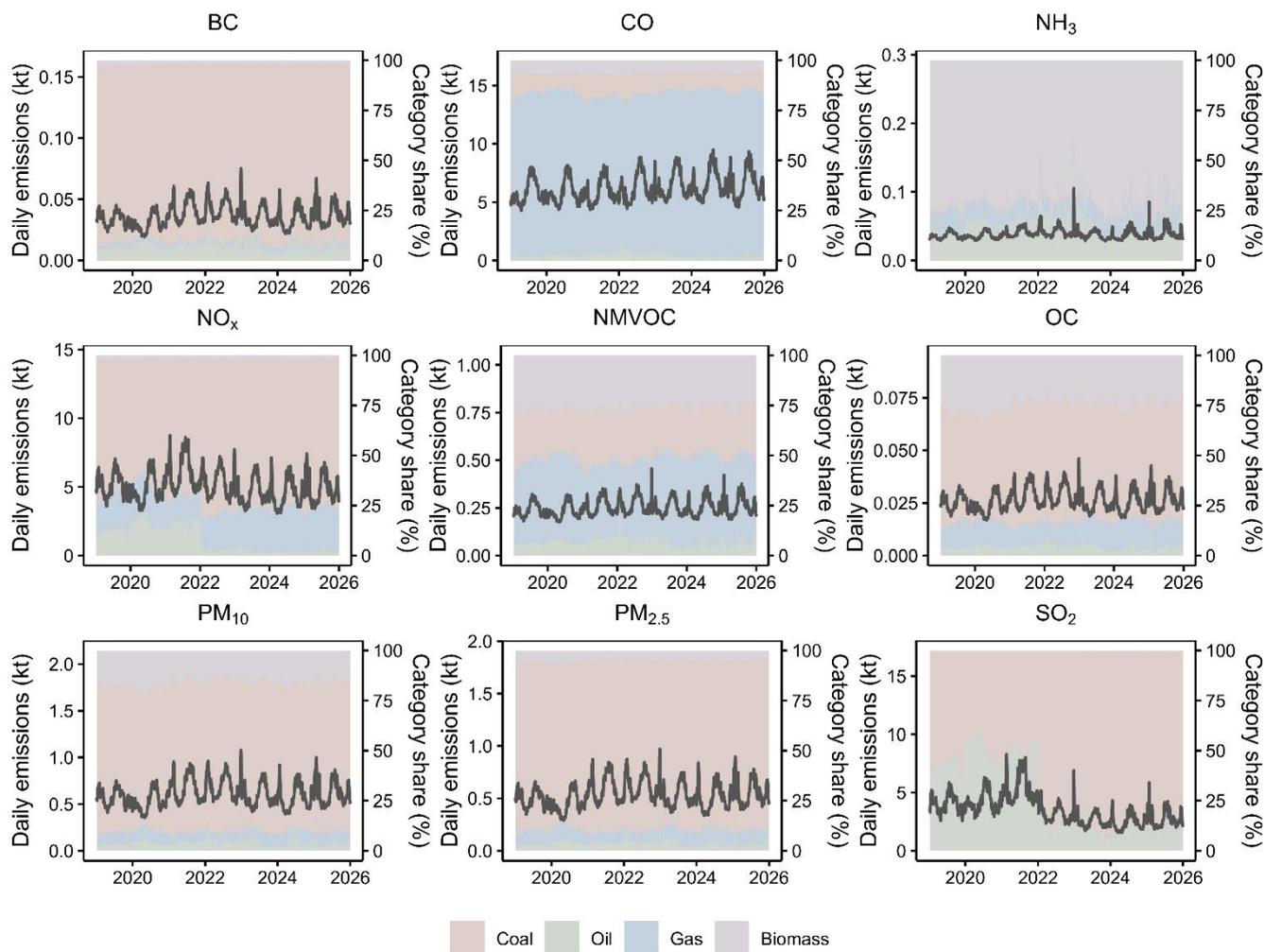

**Figure 6.** Daily total emissions and fuel-category shares of power-sector emissions in US by pollutant

By 2025, the U.S. source structure displayed a clear "coal–gas–biomass" differentiation. Coal dominated BC (87.7%), NOx (74.4%), OC (58.6%), $PM_{10}$ (75.2%), $PM_{2.5}$ (84.5%), and $SO_2$ (85.2%); natural gas dominated CO (81.6%) and NMVOC (42.5%); and $NH_3$ was mainly associated with biomass (68.4%). Over 2019–2025, coal-related emissions changed by +11.8% (BC), +8.5% (CO), +6.9% ($NH_3$), +5.2% (NOx), +5.7% (NMVOC), +11.3% (OC), +13.2% ($PM_{10}$), +13.2% ($PM_{2.5}$), and +3.9% ($SO_2$); gas-related emissions increased by 11.6%, 13.3%, 14.8%, 12.1%, 12.5%, 13.4%, 13.6%, 13.6%, and 13.0%, respectively; and biomass-related emissions were approximately stable or slightly declining, with changes of -8.4%, -2.8%, +1.5%, -4.9%, +0.6%, -8.9%, -8.1%, -2.2%, and -6.1%. Oil showed a mixed behavior, with increases in BC, CO, $NH_3$, NMVOC, OC, $PM_{10}$, and $PM_{2.5}$ but sharp declines in NOx (-80.5%) and $SO_2$ (-77.5%). Overall, the growth of U.S. emissions was associated primarily with coal and gas, while the largest reductions occurred in sulfur- and nitrogen-related emissions from oil and, to a lesser extent, from the system as a whole.

Seasonally, the United States showed a coherent annual cycle characterized by a summer peak rather than a pronounced bimodal pattern (Fig. S11). All pollutants reached their annual maxima in July, with BC, CO, $NH_3$, NOx, NMVOC, OC, $PM_{10}$, $PM_{2.5}$, and $SO_2$ peaking at 1.47, 251, 1.48, 205, 10.0, 1.05, 24.2, 21.6, and 145 kt/month, respectively. All pollutants also reached their minima in April, at 0.814, 147, 0.972, 115, 6.03, 0.617, 13.9, 12.1, and 80.4 kt/month, respectively.

### 3.2.4   Europe

Europe was the only region in which the global share of all power-sector pollutants declined consistently over 2019–2025, indicating a clear weakening of its relative contribution to the global power-plant emission budget (Fig. 7). Daily mean emissions changed from 0.0233 to 0.0234 kt/d for BC, 11.9 to 12.7 for CO, 0.0353 to 0.0334 for $NH_3$, 4.25 to 3.70 for NOx, 0.454 to 0.436 for NMVOC, 0.0232 to 0.0229 for OC, 0.524 to 0.526 for $PM_{10}$, 0.488 to 0.493 for $PM_{2.5}$, and 7.24 to 5.41 for $SO_2$. Accordingly, Europe's shares of global emissions declined from 9.96% to 8.54% for BC, 31.9% to 28.2% for CO, 10.2% to 7.99% for $NH_3$, 8.82% to 7.09% for NOx, 19.6% to 14.5% for NMVOC, 6.80% to 5.48% for OC, 9.48% to 7.78% for $PM_{10}$, 11.6% to 9.65% for $PM_{2.5}$, and 8.95% to 6.89% for $SO_2$. Therefore, Europe showed the clearest decline in relative emission importance, although the absolute decreases were most evident for NOx and $SO_2$.

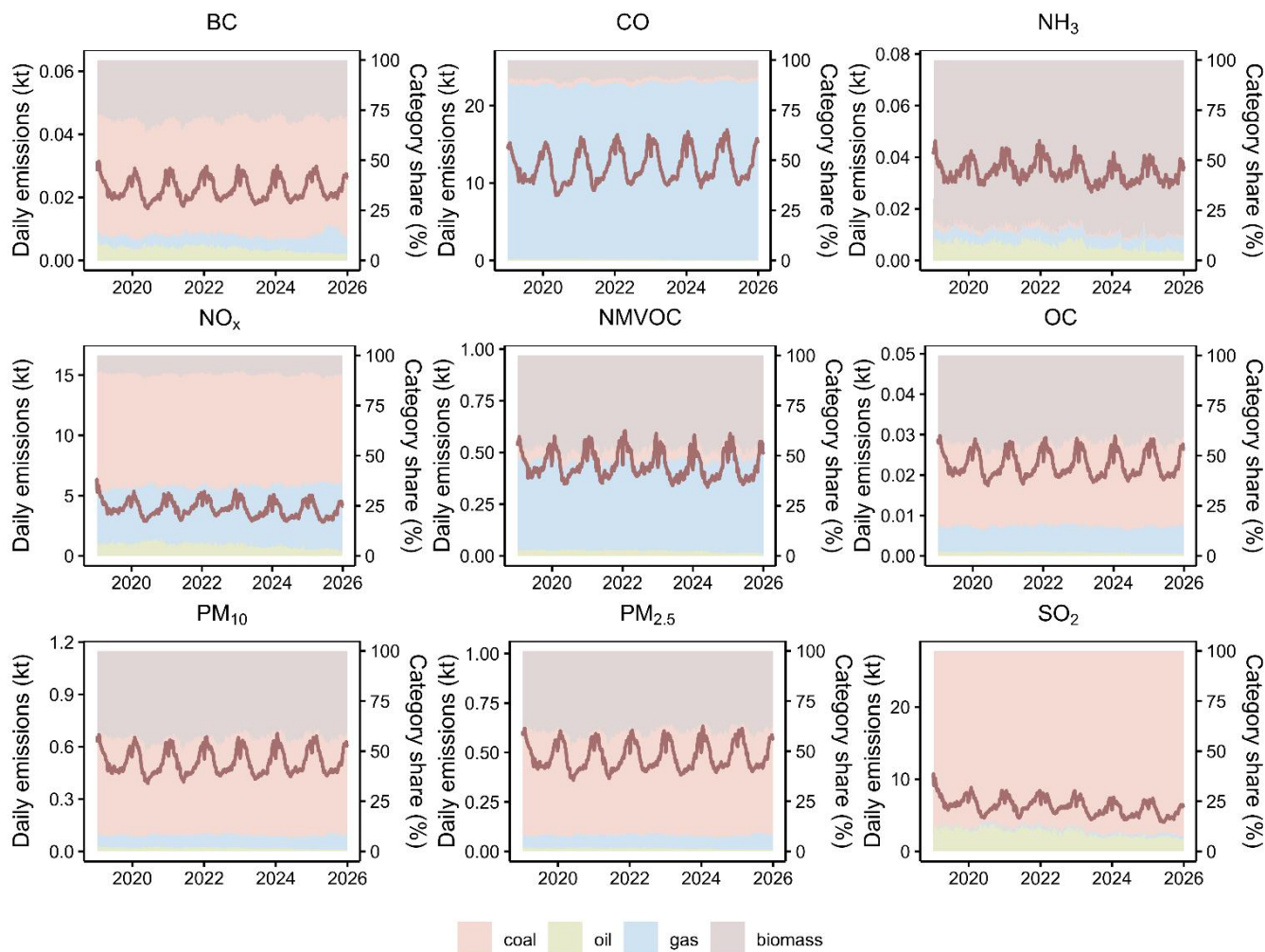

**Figure 7.** Daily total emissions and fuel-category shares of power-sector emissions in Europe by pollutant

The decline was mainly associated with reductions in coal- and oil-related emissions. Relative to 2019, coal-related emissions in 2025 changed by -1.5%, -26.9%, -61.1%, -18.7%, -27.1%, -0.5%, +0.4%, +0.8%, and -21.4% for BC, CO, NH$_3$, NOx, NMVOC, OC, PM$_{10}$, PM$_{2.5}$, and SO$_2$, respectively, while oil-related emissions decreased much more strongly, by -51.2%, -58.4%, -56.4%, -57.0%, -52.7%, -51.0%, -53.9%, -52.9%, and -58.1%. By contrast, gas-related emissions increased by +112.0%, +8.7%, +15.7%, +8.0%, +3.4%, +5.8%, +21.6%, +21.4%, and +85.6%, whereas biomass emissions were comparatively stable, changing by -1.2%, -1.8%, +1.4%, -3.0%, -4.9%, -2.1%, +0.5%, 0.0%, and -3.5%, respectively. These results indicate that Europe's power-sector transition was characterized by a marked retreat from coal and oil, partly offset by greater reliance on gas and sustained biomass use.

Europe also displayed the most coherent seasonality among all regions, with a typical winter-high and summer-low pattern (Fig. S12). All pollutants reached their maxima in January, when BC, CO, $NH_3$, NOx, NMVOC, OC, $PM_{10}$, $PM_{2.5}$, and $SO_2$ were 0.867, 470, 1.25, 152, 16.7, 0.851, 19.3, 18.1, and 240 kt/month, respectively. All pollutants reached their minima in June, declining to 0.570, 297, 0.928, 97.3, 11.3, 0.582, 13.1, 12.2, and 152 kt/month, respectively. This strong seasonal coherence suggests that Europe's power-plant emissions were closely tied to winter heating demand and lower summertime energy consumption.

## 3.3  Short-term variations of power plant emissions

Figure 8 illustrates the association between short-term temperature variability and NOx emissions. In China (2022) and the United States (2023), the minimum temperatures were observed on 24 September 2022 and 29 September 2023, reaching 19.2 and 19.7 °C, respectively, with corresponding NOx emissions of 11.8 and 4.25 kt. In contrast, the maximum temperatures occurred on 8 August 2022 and 28 July 2023, at 29.0 and 27.3 °C, respectively, and coincided with the highest NOx emissions, which reached 15.6 and 7.15 kt. Relative to the low-temperature conditions, NOx emissions increased by 32.2% in China and 68.2% in the United States under high-temperature conditions. More broadly, NOx emissions in both countries generally increased from June to August, peaked in mid-to-late summer, and then declined through September. This pattern suggests that higher temperatures are associated with enhanced NOx emissions, likely reflecting increased electricity demand for space cooling, which in turn leads to higher power generation and associated pollutant emissions.

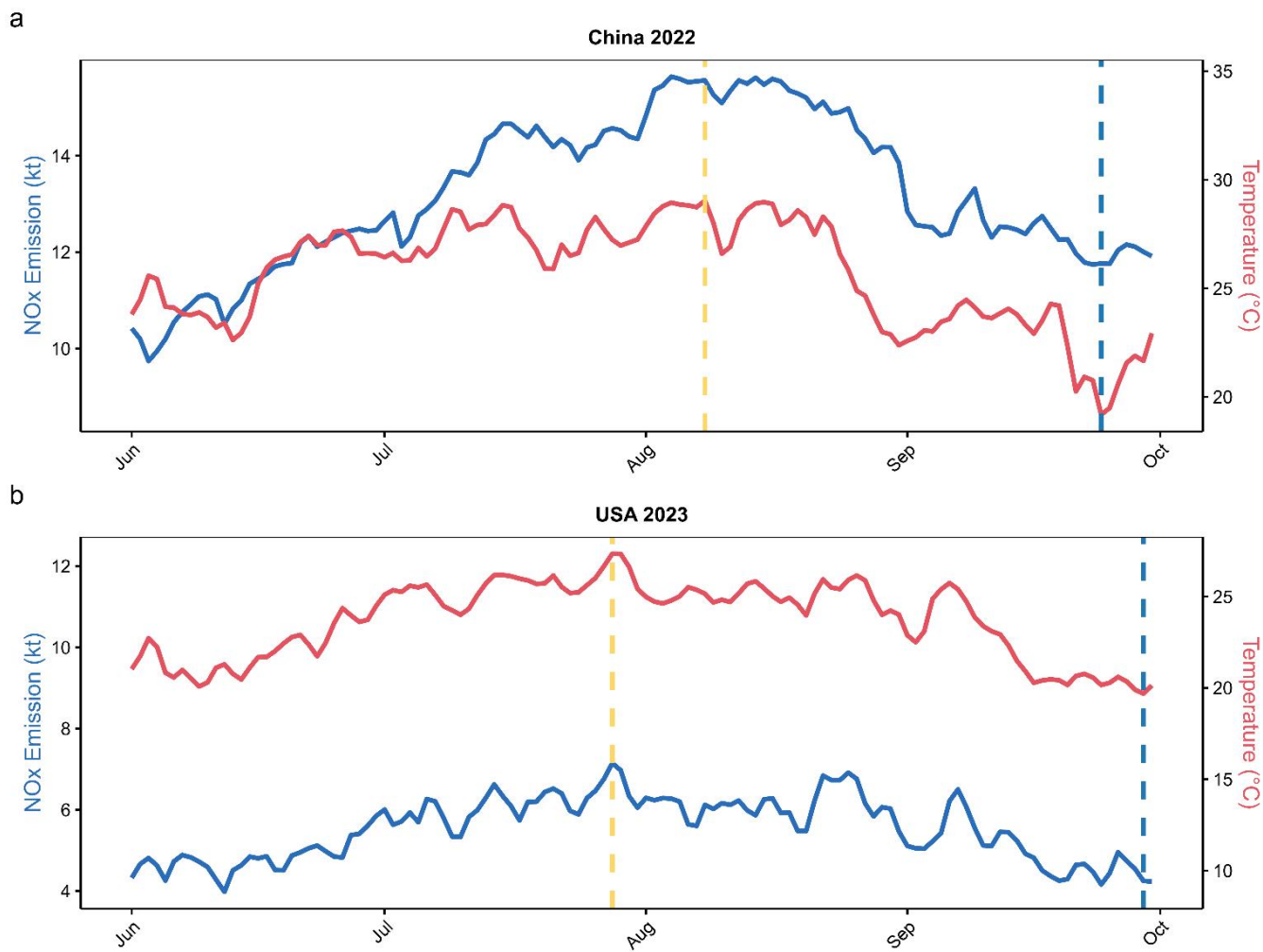

**Figure 8.** NOx emissions and temperature in (a) China and (b) the United States during June–September 2022 and June–September 2023, corresponding to reported heatwave periods. The orange and blue dashed lines indicate the maximum and minimum temperatures during these periods, respectively.

Pollutant emissions also exhibited a pronounced weekly cycle, with Sundays generally corresponding to the lowest emission levels across regions (Fig. S13). The strength of this weekly variation, however, differed substantially among regions. Europe showed the most prominent weekly cycle, with emissions remaining above the regional mean from Tuesday to Friday (by approximately 1.94%–4.88%) but decreasing markedly on Saturday and Sunday, by 6.60% and 9.94% below the mean, respectively. The United States likewise displayed a strong weekly pattern, with emissions elevated during Tuesday–Friday (by 2.52%–3.59%) and substantially reduced on Sunday and Monday, by 5.92% and 6.75% below the mean, respectively, indicating a clear weekend effect that extended into the beginning of the following week. By contrast, the weekly cycle was much weaker in China and India. In China, deviations from the mean were generally small throughout the week, with the largest reductions occurring on Saturday and Sunday, at only 0.20% and 0.92% below the mean, respectively. India also exhibited a relatively modest weekly cycle, with Sunday showing the main reduction (2.49% below the mean), whereas the remaining days stayed close to the weekly average. The differences likely reflect regional contrasts in the temporal organization of economic activity, electricity demand, and power system operation. In regions with strong weekday–weekend contrasts, emissions appear to be more strongly modulated by weekly fluctuations in industrial and commercial activity. In contrast, the relatively weak weekly cycle in China suggests that anthropogenic activity and power generation remain comparatively stable across the week, implying a less distinct separation between weekday and weekend emission behavior.

### 3.4 Comparison with bechmark dataset

Figure 9 presents a comparison between our emission estimates and the benchmark Emissions Database for Global Atmospheric Research (EDGAR) dataset. Because the most recent publicly available EDGAR emission dataset currently extends only to 2022, the comparison shown in this figure is based on the multi-year mean over the 2019-2022 period. Overall, our emission estimates are in good agreement with the EDGAR dataset, indicating that the proposed framework is able to reproduce the general magnitude and spatial characteristics of major air pollutant emissions with reasonable accuracy. Specifically, the Pearson correlation coefficients range from 0.92 to 0.99, while the mean relative deviation (MRD) ranges from 8.8% to 28.1%, demonstrating consistently strong agreement across pollutants. Among all species, the best consistency is found for $SO_2$ and $NH_3$, both of which exhibit a Pearson correlation coefficient of 0.99. Their corresponding MRDs are 11.7% for $SO_2$ and 8.8% for $NH_3$, suggesting that these two pollutants are captured particularly well by our estimates. NMVOC, CO, and OC also show strong agreement with EDGAR, with correlation coefficients of 0.98, 0.98, and 0.97, respectively, and MRDs of 12.8%, 16.0%, and 19.0%. In comparison, $PM_{10}$, BC, and NOx show relatively weaker consistency, although the agreement remains robust overall, as reflected by correlation coefficients of 0.93, 0.92, and 0.92, respectively. Their corresponding MRDs are 23.5%, 28.1%, and 19.6%. Taken together, these results suggest that our emission estimates are broadly consistent with the EDGAR benchmark, while also indicating that uncertainties remain larger for certain pollutants, particularly particulate species and nitrogen oxides.

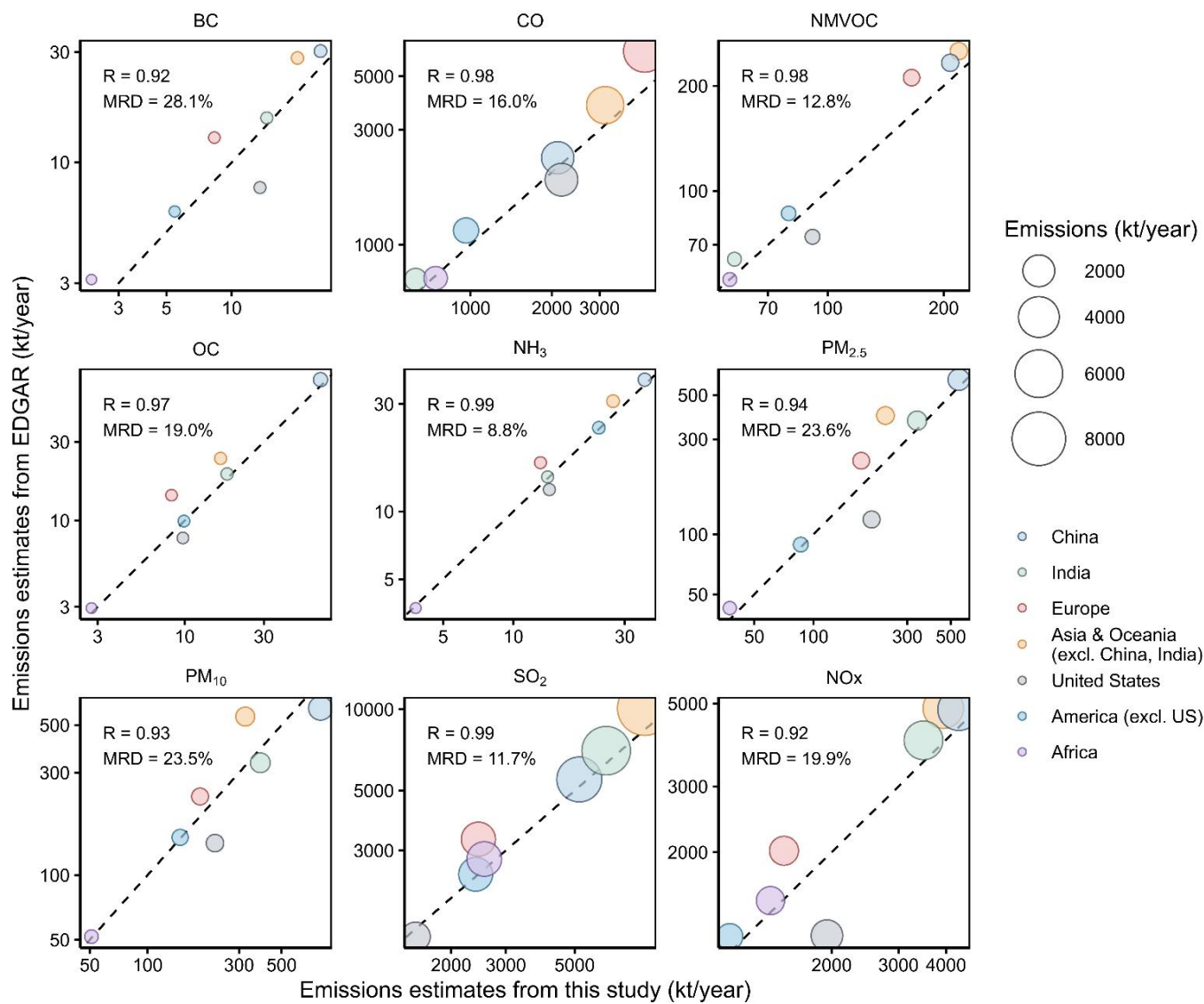

**Figure 9.** Regional comparison of our emission estimates with EDGAR. R and MRD indicate Pearson correlation coefficient and mean relative difference, respectively.

**Discussion**

A major contribution of this study is the development of a near-real-time, plant-level, daily emission dataset for the global power sector. Compared with conventional global inventories, this dataset substantially improves both temporal resolution and update timeliness, providing a more realistic representation of short-term emission variability (Janssens-Maenhout et al., 2019). Existing inventories are essential for long-term trend analysis, but their relatively coarse temporal resolution and reporting lag limit their ability to capture rapid changes in power generation and associated emissions.

The high temporal resolution of this dataset is particularly important for resolving short-term emission responses to meteorological extremes and fluctuations in electricity demand. As shown in this study, emissions, especially NOx, increase during high-temperature periods in both China and the United States. This feature is highly relevant for air quality studies because NOx, CO, and NMVOC are important ozone precursors (Li et al., 2023; Zhan et al., 2018). During extreme heat events, the coincidence of higher precursor emissions and stronger photochemical activity can aggravate ozone pollution and associated health risks (Li et al., 2023). High-resolution emission data therefore provide a more appropriate basis for assessing the air quality and health impacts of compound heat–pollution events.

The near-real-time nature of the dataset also increases its value for operational applications. Air quality forecasting systems depend strongly on timely emission inputs, but many commonly used inventories cannot reflect current emission conditions because of substantial update delays (Liu et al., 2024a; Soulie et al., 2024). By reducing this lag, the present dataset can provide more up-to-date bottom-up information for chemical transport modeling and forecast evaluation, thereby supporting more accurate air quality prediction and more responsive pollution control.

This dataset also complements existing inventories and observational approaches. Its strong agreement with EDGAR indicates that the framework captures the general magnitude and spatial pattern of power-sector emissions while providing much finer temporal detail. This makes it useful not only for emission accounting, but also for satellite inversion evaluation, model validation, and the interpretation of short-term emission anomalies (Lei et al., 2022; Lin et al., 2023; Zheng et al., 2020; Zheng et al., 2021).

Some uncertainties remain. The estimates still depend on the completeness of public generation data, the representativeness of emission factors, and the assumptions used in plant-level allocation, particularly where detailed operational information is unavailable. Future improvements should incorporate more plant-specific operating characteristics, pollution control information, and regionally refined emission factors to further enhance accuracy and consistency.

## Code and data availability

The code used to generate this dataset is available from the corresponding author upon reasonable request. The dataset generated by the current study is available on Figshare at https://doi.org/10.6084/m9.figshare.31901356 (Li et al., 2026a).

## Competing Interests

The authors declare no competing interests.

## Acknowledgements

This paper is supported by National Natural Science Foundation of China (Grant No. 72394402), and China Manned Space Program through its Space Application System (Grant No. SCP03-01-04).

## Author Contributions

**T.L.:** Writing - Original Draft, Methodology, Software, Validation, Visualization; **L.W.:** Software, Visualization, Writing - Review & Editing; **B.Z.:** Data provision, Writing - Review & Editing; **Z.L.:** Conceptualization, Supervision, Funding acquisition, Writing - Review & Editing.